\documentclass{elsart}
\newcommand{\Tr}{\mbox{Tr}}
\newcommand{\ket}[1]{\left|#1\right\rangle}
\newcommand{\norm}[1]{\Vert#1\Vert}
\begin{document}

\begin{frontmatter}



\title{Kochen-Specker Obstruction for Position and Momentum Using a Single Degree of Freedom}
\author{Wayne C. Myrvold \thanksref{thanx}}
\address{Department of Philosophy \\ University of Western Ontario \\ London, ON, Canada N6A 3K7 \\ e-mail: wmyrvold@uwo.ca}

\thanks[thanx]{The author would
like to thank Rob Clifton for pointing out a flaw in an earlier
draft of this paper, and for a suggestion regarding how to amend
the argument.}

\begin{abstract}
It is shown that, given a reasonable continuity assumption
regarding possessed values, it is possible to construct an
obstruction for any coordinate and its conjugate momentum,
demonstrating that at most one of these two quantities can have a
noncontextual value.
\end{abstract}

\begin{keyword} Kochen-Specker theorem \sep position \sep
momentum

\PACS 03.65.Bz
\end{keyword}
\end{frontmatter}

\section{Introduction} The issue of whether the uncertainty relations between position and momentum entail that these two
 quantities cannot simultaneously be `elements of reality' has long
played a central role in discussions concerning the interpretation
of quantum theory.  As is well known, the deterministic
alternative to quantum mechanics formulated by Bohm \cite{bohm}
attributes definite values of position to particles at the price
of contextualizing the momentum value---the result of a
measurement of momentum is not determined by the state of the
measured system alone but depends crucially on the details of the
experimental arrangement.  The Bell-Kochen-Specker theorem
\cite{bks} shows that it is not possible to attribute simultaneous
noncontextual values to all physical quantities.  It seems
eminently plausible that an argument analogous to the
Bell-Kochen-Specker argument can show that, as a consequence of
the canonical commutation relations, at most one of a pair of
canonically conjugate observables can have a noncontextual value.
Proof of such an assertion has been surprisingly long in coming.
Clifton \cite{clif}, building on the work of Peres and Mermin
\cite{peres}, exhibited a Kochen-Specker obstruction utilizing two
degrees of freedom. This shows that it is not possible for two
independent coordinates $q_1$, $q_2$, and their conjugate momenta
$p_1$, $p_2$ to all have noncontextual values. In this paper this
conclusion is extended to a single degree of freedom; of any pair
of canonical conjugates $q$, $p$, at most one can have a
noncontextual value.

The  obstruction constructed here differs from previous
obstructions in one important respect, in that the usual
assumptions about measured values will be supplemented with a
continuity assumption. Kochen-Specker obstructions usually assume
the Product Rule:  if observables associated with commuting
operators $\hat{A}$, $\hat{B}$ have definite values $v[\hat{A}]$,
$v[\hat{B}]$, then the product operator $\hat{A}\hat{B}$ has
associated with it the definite value $v[\hat{A}\hat{B}] =
v[\hat{A}]\: v[\hat{B}]$.  We will extend a version of this to
noncommuting operators that correspond to observables that are
\emph{approximately} co-measurable, in a sense that will be
explained in section \ref{cont}.

\section{The Peres/Mermin and Clifton obstructions} The following is a generalization
of  the simple Kochen-Specker obstruction constructed by Mermin on
the basis of work by Peres \cite{peres}. Take operators
$\hat{A}_1$, $\hat{A}_2$, $\hat{B}_1$, $\hat{B}_2$, such that none
of them have zero as an eigenvalue, satisfying the commutation
relations,
\begin{equation}\label{comm}
[\hat{A}_1, \hat{A}_2] = [\hat{A}_1, \hat{B}_2]= [\hat{B}_1,
\hat{A}_2] = [\hat{B}_1, \hat{B}_2] = 0,
\end{equation}
and the anticommutation relations
\begin{equation}\label{anti}
\hat{A}_i \hat{B}_i = - \hat{B}_i \hat{A}_i,\quad i = 1,2.
\end{equation}
(The specific example used by Mermin, for a pair of spin-$\half$
particles, is $\hat{A}_i = \hat{\sigma}_{i x}$, $\hat{B}_i =
\hat{\sigma}_{i y}$). Suppose that associated with $\hat{A}_1$,
$\hat{A}_2$, $\hat{B}_1$, $\hat{B}_2$ are nonzero definite values
$v[\hat{A}_1]$, $v[\hat{A}_2]$, $v[\hat{B}_1]$, $v[\hat{B}_2]$,
respectively. As a consequence of the commutation relations
(\ref{comm}) and the Product Rule, we have
\begin{eqnarray} \nonumber
{v[\hat{A}_1 \hat{A}_2] = v[\hat{A}_1] \: v[\hat{A}_2]}
\\
{v[\hat{B}_1 \hat{B}_2] = v[\hat{B}_1] \: v[ \hat{B}_2].}
\end{eqnarray}
Moreover, $\hat{A}_1 \hat{A}_2$ commutes with $\hat{B}_1
\hat{B}_2$, and so
\begin{eqnarray}\label{a}
\nonumber v[\hat{A}_1 \hat{A}_2 \hat{B}_1 \hat{B}_2] &=&
v[\hat{A}_1 \hat{A}_2] \: v[\hat{B}_1 \hat{B}_2] \\ &=&
v[\hat{A}_1] \: v[ \hat{A}_2] \: v[ \hat{B}_1] \: v[\hat{B}_2].
\end{eqnarray}
As a further consequence of the commutation relations and the
Product Rule, we also have
\begin{eqnarray} \nonumber
v[\hat{A}_1 \hat{B}_2] &=& v[\hat{A}_1] \: v[\hat{B}_2]
\\
v[\hat{A}_2 \hat{B}_1] &=& v[\hat{A}_2] \: v[\hat{B}_1].
\end{eqnarray}
Since $\hat{A}_1 \hat{B}_2$ commutes with $\hat{A}_2 \hat{B}_1$,
\begin{eqnarray}\label{b}
\nonumber v[\hat{A}_1 \hat{B}_2 \hat{A}_2 \hat{B}_1] &=&
v[\hat{A}_1 \hat{B}_2] \: v[ \hat{A}_2 \hat{B}_1] \\ &=&
v[\hat{A}_1] \: v[\hat{B}_2] \: v[\hat{A}_2] \: v[\hat{B}_1].
\end{eqnarray}
Comparison of (\ref{a}) and (\ref{b}) yields
\begin{equation}\label{c}
v[\hat{A}_1 \hat{A}_2 \hat{B}_1 \hat{B}_2] = v[\hat{A}_1 \hat{B}_2
\hat{A}_2 \hat{B}_1].
\end{equation}
However, we also have
\begin{equation}
\hat{A}_1 \hat{B}_2 \hat{A}_2 \hat{B}_1 = -\hat{A}_1 \hat{A}_2
\hat{B}_2 \hat{B}_1 = -\hat{A}_1 \hat{A}_2 \hat{B}_1 \hat{B}_2.
\end{equation}
Consequently,
\begin{equation}\label{d}
v[\hat{A}_1 \hat{A}_2 \hat{B}_1 \hat{B}_2] = -v[\hat{A}_1
\hat{B}_2 \hat{A}_2 \hat{B}_1],
\end{equation}
and hence
\begin{equation}
v[\hat{A}_1] \: v[\hat{B}_2] \: v[\hat{A}_2] \: v[\hat{B}_1] = -
v[\hat{A}_1] \: v[\hat{B}_2] \: v[\hat{A}_2] \: v[\hat{B}_1],
\end{equation}
contradicting the assumption that $v[\hat{A}_1]$, $v[\hat{A}_2]$,
$v[\hat{B}_1]$, $v[\hat{B}_2]$ are all nonzero.

Clifton \cite{clif} used the Weyl form of the canonical
commutation relations,
\begin{eqnarray}\label{weyl1}
\nonumber [e^{- i a \hat{q}_i / \hbar}, \, e^{-i b \hat{q}_j /
\hbar}] &=& [e^{- i a \hat{p}_i / \hbar}, \, e^{- i b \hat{p}_j /
\hbar}]
\\ &=& [e^{- i a \hat{q}_i / \hbar}, \,  e^{- i b \hat{p}_j / \hbar}] = 0 ,
\; \; i \neq j
\end{eqnarray}
\begin{equation}\label{weyl2}
e^{- i a \hat{q}_i / \hbar} \: e^{- i b \hat{p}_i / \hbar} = e^{-i
a b / \hbar} e^{- i b \hat{p}_i / \hbar}\: e^{- i a \hat{q}_i /
\hbar}
\end{equation}
to construct non-Hermitian unitary operators satisfying
(\ref{comm}) and (\ref{anti}).  Clifton's  obstruction is,
therefore, not obtained directly for real-valued observables but
via a detour through the complex plane; this is supplemented by an
argument that an obstruction in the complex plane entails one on
the real line. It is not difficult to exhibit an obstruction
directly in terms of Hermitian operators and real-valued
observables.  Let $a_1$, $b_1$, $a_2$, $b_2$ be real numbers such
that
\begin{eqnarray}
\nonumber a_1 \, b_1 = (2m + 1) \pi \hbar
\\
a_2 \, b_2 = (2n + 1) \pi \hbar
\end{eqnarray}
for some integers $m$, $n$.  Let $\hat{q}_1$, $\hat{q}_2$ be the
operators corresponding to two independent coordinates, and let
$\hat{p}_1$, $\hat{p}_2$ be the conjugate momenta operators.
Define the operators
\begin{eqnarray}
\nonumber {\hat{A}_i = \cos(a_i \hat{q}_i / \hbar)}
\\
{\hat{B}_i = \cos(b_i \hat{p}_i / \hbar).}
\end{eqnarray}
It follows from the Weyl commutation relations (\ref{weyl1},
\ref{weyl2}) that these operators satisfy (\ref{comm}) and
(\ref{anti}), and hence we have the desired obstruction.

\section{The Continuity Assumption}\label{cont} To exhibit a Kochen-Specker
obstruction using only a single coordinate $q$ and its conjugate
momentum, we will construct operators satisfying the
anticommutation relations (\ref{anti}).  However, we will not be
able to satisfy at the same time the commutation relations
 (\ref{comm}). The rationale for the usual Product
Rule is that commuting operators correspond to observables that
are co-measurable: an ideal measurement of one quantity does not
disturb the value of the other.  Our obstruction will involve
noncommuting operators that are \emph{approximately}
co-measurable. Let us consider the disruption of the statistical
distribution of one observable by a measurement of another. Let
$\hat{A}$ be a compact operator, and let $\{\hat{P}^A_i\}$ be its
spectral projections. Suppose that a system in initial state
$\hat{\rho}$ is subjected to one of two procedures--- either
$\hat{B}$ alone is measured, or $\hat{B}$ is measured subsequent
to a prior ideal measurement of $\hat{A}$.  In the former case,
the expectation value of the result of the $\hat{B}$-measurement
is
\begin{equation}
\langle \hat{B} \rangle _{\hat{\rho}} =  \Tr(\hat{\rho} \hat{B}).
\end{equation}
If, however, the $\hat{A}$-measurement is performed first, the
expectation value for the result of the $\hat{B}$-measurement is
\begin{equation}
\langle \hat{B}' \rangle _{\hat{\rho}} = \Tr(\sum_i \hat{P}^A_i
\hat{\rho} \hat{P}^A_i \hat{B}) =  \Tr( \hat{\rho} \sum_i
\hat{P}^A_i \hat{B} \hat{P}^A_i).
\end{equation}
The difference between the two expectation values is
\begin{equation}
\langle \hat{B}' \rangle _{\hat{\rho}} - \langle \hat{B} \rangle
_{\hat{\rho}} = \Tr(\hat{\rho}\; (\sum_i \hat{P}^A_i \hat{B}
\hat{P}_i - B )) = - \Tr(\hat{\rho} \; \sum_i (I-\hat{P}^A_i)
\hat{B} \hat{P}^A_i).
\end{equation}
Define the disturbance of $\hat{B}$ by $\hat{A}$,
\begin{equation}
\Delta(\hat{B};\hat{A}) = - \sum_i (I-\hat{P}^A_i) \hat{B}
\hat{P}^A_i
\end{equation}
If $\hat{A}$ and $\hat{B}$ commute, the disturbance
$\Delta(\hat{B};\hat{A})$ is the zero operator. If $\hat{A}$ and
$\hat{B}$ don't commute but $\Delta(\hat{B};\hat{A})\ket{\psi} =
0$, a measurement of $\hat{A}$ leaves the expectation value of
$\hat{B}$ unchanged but may alter the statistical distribution of
these values about the expectation value.  What we want to demand,
in order to regard the value of $\hat{B}$ as minimally disturbed,
is that the probability distribution for the outcome of a
$\hat{B}$-measurement be minimally disrupted.  If, in addition, a
measurement of $\hat{A}$ only minimally disturbs the statistics
regarding the product of $\hat{A}$ and $\hat{B}$, we ought to
ascribe, at least with high probability, a definite value to the
product of $\hat{A}$ and $\hat{B}$ that is close to the product of
the values of $\hat{A}$ and $\hat{B}$.  Of course, unless
$\hat{A}$ and $\hat{B}$ commute, the product operator
$\hat{A}\hat{B}$ will not be Hermitian, and hence not correspond
to any observable; we will attribute a definite value instead to
the symmetrized product,
\begin{equation}
\hat{A} \circ \hat{B} = \half (\hat{A} \hat{B} + \hat{B} \hat{A}).
\end{equation}
We will want to consider  sequences of operators $\{\hat{A}_n\}$,
$\{\hat{B}_n\}$ such that, for any state, the disturbance of the
the statistics for $\hat{B}_n$ can be made as small as one likes
by taking $n$ sufficiently large. The statistical distribution of
the results of $\hat{B}$-measurements will be  minimally altered
if the expectation value of $\hat{B}^k$ is minimally altered for
all $k$ less than some sufficiently large $K$; that is, we can
approximate any distribution by approximately recovering, for
sufficiently high $K$, the first $K$ moments of the distribution.

The assumption about definite values on which  our obstruction
will be based is the following.
\begin{quote} Let $\{\hat{A}_n\}$,  $\{\hat{B}_n\}$ be
sequences of operators such that, for every natural number $k$ and
every  vector $\ket{\psi}$,  $\norm{\Delta(\hat{B}_n^k ;\hat{A}_n)
\ket{\psi}}$ and $\norm{\Delta((\hat{A}_n \circ
\hat{B}_n)^k;\hat{A}_n) \ket{\psi}}$  converge to zero as $n
\rightarrow \infty$. If $\hat{A}_n$ and $\hat{B}_n$ have definite
values $v[\hat{A}_n]$ and $v[\hat{B}_n]$, then, for any state
$\hat{\rho}$, and any $\epsilon, \delta > 0$, there exists $N$
such that for all $n > N$ the probability is greater than $1 -
\delta$ that the symmetrized product $\hat{A}_n \circ \hat{B}_n$
has a definite value satisfying $$\left|v[\hat{A}_n \circ
\hat{B}_n] - v[\hat{A}_n]v[\hat{B}_n]\right| < \epsilon.$$
\end{quote}

A sequence of operators $\{\hat{\Delta}_n\}$ is said to
\emph{strongly converge} to a limit $\hat{\Delta}$ if and only if,
for any vector $\ket{\psi}$, $\norm{(\hat{\Delta}_n - \hat{\Delta}
) \ket{\psi}} \rightarrow 0$ as $n \rightarrow \infty$. Therefore,
the condition in the first sentence of the above rule is the
condition that the sequences of operators $\{\Delta(\hat{B}_n^k
;\hat{A}_n)\}$ and $\{\Delta((\hat{A}_n \circ
\hat{B}_n)^k;\hat{A}_n)\}$ strongly converge to zero.

A construction invoking this rule, which we will call the
$\epsilon$-Product Rule, will be called an $\epsilon$-obstruction.
Note that the usual Product Rule is entailed as a special case by
the $\epsilon$-Product Rule. It is also worth noting that the
$\epsilon$-Product Rule is satisfied by the simple noncontextual
hidden-variable theory constructed by Bell \cite{bks} for a system
consisting of a single spin-$\half$ particle.

The $\epsilon$-obstruction constructed in section \ref{ob} will
not be a state-independent one, as we will, in general, have to
choose  for different states different values of $n$ to obtain a
set of operators composing the obstruction.  What will be shown is
that, for any state,  there is a set of operators that cannot
consistently be assigned values in accord with the
$\epsilon$-Product rule but which are required by that rule to
have definite values if all functions of $\hat{q}$ and all
functions of $\hat{p}$ are ascribed definite values. The
conclusion is that no hidden-variables theory satisfying the
$\epsilon$-Product Rule can attribute noncontextual values to both
of $\hat{q}$ and $\hat{p}$, in any state.

\section{The Obstruction}\label{ob}  Suppose there were numbers $a_1$,
$a_2$, $b_1$, $b_2$ such that
\begin{eqnarray}\label{fantasy}
\nonumber a_1 \, b_1 &=& (2m+1) \pi \hbar \\  \nonumber a_1 \, b_2
&=& 2k \, \pi \hbar
\\ \nonumber a_2 \, b_1 &=& 2l \, \pi \hbar \\  a_2 \, b_2 &=&   (2n +1) \pi
\hbar
\end{eqnarray}
for some integers $k$, $l$, $m$, $n$.  Then we could readily
construct an obstruction for a single degree of freedom by
defining the operators,
\begin{eqnarray}
\nonumber \hat{A}_i = \cos(a_i \hat{q} / \hbar)
\\
\hat{B}_i = \cos(b_i \hat{p} / \hbar).
\end{eqnarray}
As can easily be verified, such operators would satisfy the
commutation /anticommutation  relations (\ref{comm}, \ref{anti}).
There are no integers satisfying equations (\ref{fantasy}),
however. To see this, note that this would require the product
$a_1 b_1 a_2 b_2$ to be both an even and an odd multiple of $\pi^2
\hbar^2$. What we will do instead is satisfy these equations
approximately.  Define
\begin{eqnarray}
\nonumber a_{1n} &=& (2n + 1)\, a
\\
\nonumber a_{2n} &=& (2 + 1 / 2n )\, a
\\
\nonumber b_1 &=& \pi \hbar / a
\\
b_{2n} &=&  2 n  \;  \pi \hbar / a,
\end{eqnarray}
where $a$ is an arbitrary constant.  Then we have
\begin{eqnarray}
\nonumber a_{1 n} \: b_1 &=& (2n + 1) \, \pi \hbar
\\ \nonumber
a_{1 n} \: b_{2 n} &=& 2 n (2n+1) \, \pi \hbar
\\ \nonumber
a_{2 n} \: b_1 &=& (2 + 1/2n)\, \pi \hbar
\\
a_{2 n} \: b_{2n} &=& (4n + 1)\, \pi \hbar.
\end{eqnarray}
Define the  projection operators
\begin{eqnarray} \nonumber
\hat{E}^{+}_{i\, n} &=& \hat{E}[\: \cos(a_{i n} \: \hat{q} /
\hbar) \geq 0 \:]  \\ \hat{E}^{-}_{i \, n} &=& \hat{E}[\:\cos(a_{i
n} \: \hat{q} / \hbar) < 0 \:], \qquad i = 1,2.
\end{eqnarray}
where $\hat{E}[\cdot]$ denotes the spectral projection onto the
specified subspace.  We will also have occasion to consider
translations of these operators,
\begin{eqnarray} \nonumber
\hat{E}^{+}_{i \,n}(\epsilon) &=& \hat{E}[\: \cos(a_{i n} \:
\hat{q} / \hbar  +  \epsilon \hat{I}) \geq 0 \: ]  \\
\hat{E}^{-}_{i \, n}(\epsilon) &=& \hat{E}[\: \cos(a_{i n} \:
\hat{q} / \hbar + \epsilon \hat{I} ) < 0 \:].
\end{eqnarray}
Now define the operators
\begin{eqnarray}
\nonumber \hat{A}_{i  n} &=& \hat{E}^{+}_{i \, n} - \hat{E}^{-}_{i
\, n}
\\ \nonumber
\hat{B}_1 &=& \cos \left(b_1 \, \hat{p} / \hbar \right)
\\
\hat{B}_{2 n} &=& \cos \left(b_{2 n} \, \hat{p} / \hbar\right).
\end{eqnarray}
We will repeatedly use the relation,
\begin{equation}
f(\hat{q}) \: e^{i b \hat{p}/\hbar}  =  e^{i b \hat{p}/\hbar} \:
f(\hat{q} - b \, \hat{I} ).
\end{equation}
If $f$ is a periodic function of period $b$,
\begin{equation}
\cos(b \, \hat{p} / \hbar) f(\hat{q}) =   f(\hat{q}) \cos(b \,
\hat{p} / \hbar),
\end{equation}
and, if $f$ changes sign under a translation by an amount $b$,
\begin{equation}
\cos(b \, \hat{p} / \hbar) f(\hat{q}) =    - f(\hat{q}) \cos(b \,
\hat{p} / \hbar).
\end{equation}
$\hat{A}_{i n}$ is a periodic function of $\hat{q}$ with period $2
\pi \hbar / a_{i n}$; it changes sign under a translation by an
odd multiple of $\pi \hbar / a_{i n}$.  We therefore have the
commutation relations
\begin{equation}\label{commn}
[\hat{A}_{1 n},\; \hat{A}_{2 n}] = [\hat{A}_{1 n}, \; \hat{B}_{2
n}] = [\hat{B}_1, \; \hat{B}_{2 n}] = 0,
\end{equation}
and the anticommutation relations
\begin{eqnarray}\label{antin}
\nonumber \hat{A}_{1 n} \hat{B}_1 &=& -\hat{B}_1 \hat{A}_{1 n}
\\
\hat{A}_{2 n} \hat{B}_{2 n} &=& -\hat{B}_{2 n} \hat{A}_{2 n}.
\end{eqnarray}
The operators $\hat{E}^+_{2 n}$, $\hat{E}^-_{2 n}$, $\hat{A}_{2 n}
$ do not, for any $n$, commute with $\hat{B}_1$. What we have
instead is
\begin{eqnarray}
\nonumber \hat{E}^{\pm}_{2 n} \hat{B}_1  &=& \half \, e^{i b_1
\hat{p} / \hbar} \: \hat{E}^{\pm}_{2 n}(-\epsilon_n) + \half \,
e^{- i b_1 \hat{p} / \hbar} \: \hat{E}^{\pm}_{2 n}(+\epsilon_n)
\\ \nonumber
\hat{A}_{2 n} \hat{B}_1  &=& \half \, e^{i b_1 \hat{p} / \hbar} \:
(\hat{E}^{+}_{2 n}(-\epsilon_n) - \hat{E}^{-}_{2 n}(-\epsilon_n))
\\& \qquad & \qquad + \half \, e^{- i b_1 \hat{p} / \hbar} \: (\hat{E}^{+}_{2
n}(+\epsilon_n) - \hat{E}^{-}_{2 n}(+\epsilon_n)),
\end{eqnarray}
where $\epsilon_n = \pi/2n.$

Let us consider the disruption of the statistics for $\hat{B}_1$
by a measurement of $\hat{A}_{2 n}$.  We have
\begin{equation}\label{del}
\Delta(\hat{B}_1 ; \hat{A}_{2 n}) = -\hat{E}^{-}_{2 n } \hat{B}_1
\hat{E}^{+}_{2 n} - \hat{E}^{+}_{2 n} \hat{B}_1 \hat{E}^{-}_{2 n}.
\end{equation}
The first term of this is
\begin{eqnarray}
- \hat{E}^{-}_{2 n } \hat{B}_1 \hat{E}^{+}_{2 n} = - \half \, e^{i
b_1 \hat{p} / \hbar} \: \hat{E}^{-}_{2
n}(-\epsilon_n)\hat{E}^{+}_{2 n} - \half \, e^{- i b_1 \hat{p} /
\hbar}\hat{E}^{-}_{2 n}(+ \epsilon_n)\hat{E}^{+}_{2 n}.
\end{eqnarray}
The operator $\hat{E}^{-}_{2 n}(-\epsilon_n)\hat{E}^{+}_{2 n}$ is
the projection onto the part of the spectrum of $\hat{q}$ such
that $\cos(a_{2 n} \: {q}) \geq 0$ but $\cos(a_{2 n} \, {q} -
\epsilon_n) < 0$. For any state $\hat{\rho}$, the measure of this
part of the spectrum in any interval $[-M, M]$ will be small if
$n$ is large, and will converge to zero as $n \rightarrow \infty$.
Hence, for any state with compact support in $\hat{q}$ ,
$\norm{\hat{E}^{-}_{2 n}(-\epsilon_n)\hat{E}^{+}_{2 n} \,
\hat{\rho} \:} \rightarrow 0$ as $n \rightarrow \infty$ (recall
that $\hat{q}$ has no eigenvalues), and, since such states are
norm-dense in the set of all states, $\norm{\hat{E}^{-}_{2
n}(-\epsilon_n)\hat{E}^{+}_{2 n} \, \hat{\rho} \:} \rightarrow 0$
as $n \rightarrow \infty$  for any state $\rho$.  Analogous
considerations apply to the other terms in the expansion of
$\Delta( \hat{B}_1, \hat{A}_{2 n})$. We conclude that the
$\Delta(\hat{B}_1 ; \hat{A}_{2 n})$ strongly converges to zero as
$n \rightarrow \infty$. A similar argument shows that, for any
$k$, $\Delta(\hat{B}^k_1 ; \hat{A}_{2 n})$ strongly converges to
zero as $n \rightarrow \infty$.

We are now ready to begin the construction of our
$\epsilon$-obstruction.  Suppose that associated with $\hat{A}_{1
n} $, $\hat{A}_{2 n}$, $\hat{B}_1$, $\hat{B}_{2 n}$ are definite
values $v[\hat{A}_{1 n}]$, $v[\hat{A}_{2 n}]$, $v[\hat{B}_1]$,
$v[\hat{B}_{2 n}]$.  Since the spectrum of each $\hat{A}_{1 n}$,
$\hat{A}_{2 n}$ is contained in $\{-1, 1\}$, the corresponding
values will be nonzero.  Moreover, we can, without loss of
generality (or rather, a loss of measure zero)  assume
$v[\hat{B}_1]$, $v[\hat{B}_{2 n}]$ to be nonzero also, since, for
any state, the probability is zero that a measurement of either
will yield a result \emph{exactly} equal to zero.  Choose some
positive $\delta < \half$.

Since $\hat{A}_{1 n}$ commutes with $\hat{B}_{2 n}$, the Product
Rule requires that associated with the product operator
$\hat{A}_{1 n} \hat{B}_{2 n}$ is the definite value
\begin{equation}
v[\hat{A}_{1 n} \hat{B}_{2 n}] = v[\hat{A}_{1 n}]\, v[\hat{B}_{2
n}].
\end{equation}
As mentioned, $\hat{A}_{2 n}$ does not commute with $\hat{B}_1$.
However, as we have seen, for any state $\hat{\rho}$, the
disturbance of the statistics of $\hat{B}$-measurements by a
measurement of $\hat{A}_{2 n}$  can be made arbitrarily small by
taking $n$ sufficiently large.  Moreover, it is easy to verify
that $\hat{A}_{2 n}$ commutes with the symmetrized product
$\hat{B}_1 \circ \hat{A}_{2 n}$, since, for any $\hat{A}$,
$\hat{B}$,
\begin{equation}
[\hat{A}, \: \hat{A} \circ \hat{B}] = \half \: [\hat{A}^2, \:
\hat{B}],
\end{equation}
and the square of $\hat{A}_{2 n}$ is the identity operator.
Therefore, measurements of $\hat{A}_{ 2n}$ do not disrupt that
statistics of measurements of $\hat{B}_1 \circ \hat{A}_{2 n}$ and,
for sufficiently large $n$, only minimally disrupt the statistics
of $\hat{B}_1$-measurements. The $\epsilon$-Product Rule dictates
that, for any $\epsilon > 0$,  for sufficiently large $n$ there is
a probability greater than $1-\delta$ that the symmetrized product
$\hat{B}_1 \circ \hat{A}_{2 n}$ has a definite value satisfying
\begin{equation}
\left|v[\hat{B}_1 \circ \hat{A}_{2 n}] - v[\hat{B}_1] \,
v[\hat{A}_{2 n}]\right| < \epsilon.
\end{equation}

$\hat{B}_1 \circ \hat{A}_{2 n}$ commutes with $\hat{B}_{2 n} \,
\hat{A}_{1 n}$.  Therefore, if $\hat{B}_1 \circ \hat{A}_{2 n}$ has
a definite value $v[\hat{B}_1 \circ \hat{A}_{2 n}]$, then
\begin{eqnarray}
\nonumber v[(\hat{B}_1 \circ \hat{A}_{2 n})\; \hat{B}_{2 n}
\hat{A}_{1 n}] &=& v[\hat{B}_1 \circ \hat{A}_{2 n}] \,
v[\hat{B}_{2 n} \hat{A}_{1 n}] \\   &=& v[\hat{B}_1 \circ
\hat{A}_{2 n}]\, v[\hat{B}_{2 n}] \, v[\hat{A}_{1 n}],
\end{eqnarray}
and hence
\begin{eqnarray}
\nonumber v[(\hat{B}_1 & \circ & \hat{A}_{2 n})\; \hat{B}_{2 n}
\hat{A}_{1 n}] - v[\hat{B}_1]\, v[\hat{A}_{2 n}]\, v[\hat{B}_{2
n}]\, v[\hat{A}_{1 n}] \\ &=& (v[\hat{B}_1 \circ \hat{A}_{2 n}] -
v[\hat{B}_1 \circ \hat{A}_{2 n}]) \, v[\hat{B}_{2 n}] \,
v[\hat{A}_{1 n}]
\end{eqnarray}

We thus conclude that, for any $\epsilon$, for sufficiently large
$n$ the probability is greater than $1 - \delta$ that $(\hat{B}_1
\circ \hat{A}^{(n)}_2)\; \hat{B}^{(n)}_2 \hat{A}^{(n)}_1$ has a
definite value satisfying
\begin{eqnarray}\label{aa}
\nonumber \left|v[(\hat{B}_1  \circ  \hat{A}_{2 n})\; \hat{B}_{2
n} \hat{A}_{1 n}]\right. - \left. v[\hat{B}_1]\, v[\hat{A}_{2
n}]\, v[\hat{B}_{2 n}]\, v[\hat{A}_{1 n}]\right|\\ = \nonumber
\left|v[\hat{B}_1 \circ \hat{A}_{2 n}] \right. - \left.
v[\hat{B}_1 \circ \hat{A}_{2 n}]\right| \left|v[\hat{B}_{2 n}] \,
v[\hat{A}_{1 n}]\right|
\\ < \quad \epsilon \; \left|v[\hat{B}_{2 n}] \,
v[\hat{A}_{1 n}]\right| \leq \epsilon,
\end{eqnarray}
where the last step is justified by the fact that the spectra of
$\hat{B}_{2 n}$ and $\hat{A}_{1 n}$ are contained in $[-1,1]$.

Since $\hat{A}_{1 n}$ and $\hat{A}_{2 n}$ commute, the Product
Rule gives,
\begin{equation}
v[\hat{A}_{1 n} \hat{A}_{2 n}] = v[\hat{A}_{1 n}] \, v[ \hat{A}_{2
n}].
\end{equation}
Similarly,
\begin{equation}
 v[\hat{B}_1 \hat{B}_{2 n}]= v[\hat{B}_1] \, v[\hat{B}_{2 n}].
\end{equation}
The operators $\hat{A}_{1 n} \hat{A}_{2 n}$ do not commute with
$\hat{B}_1 \hat{B}_{2 n}$.  However, an argument analogous to that
used above for the case of  $\hat{A}^{(n)}_2$ and $\hat{B}_1$,
shows that, for any $k$, $\Delta((\hat{B}_1 \hat{B}_{2n})^k ;
\hat{A}_{1 n} \hat{A}_{2 n}) $ converges strongly to zero as $n
\rightarrow \infty$ (the details of this argument present no novel
features and will not be rehearsed here).  Moreover, $\hat{A}_{1
n} \hat{A}_{2 n}$ commutes with $(\hat{B}_1 \hat{B}_{2 n}) \circ
(\hat{A}_{1 n} \hat{A}_{2 n}) $.  Therefore, for any $\epsilon >
0$, for sufficiently large $n$ the probability is greater than $1-
\delta$ that $(\hat{B}_1 \hat{B}_{2 n}) \circ (\hat{A}_{1 n}
\hat{A}_{2 n})$ has a definite value satisfying
\begin{equation}
\left|v[(\hat{B}_1 \hat{B}_{2 n}) \circ (\hat{A}_{1 n} \hat{A}_{2
n})] - v[\hat{B}_1 \hat{B}_{2 n}] \, v[\hat{A}_{1 n} \hat{A}_{2
n})\right| < \epsilon,
\end{equation}
and hence
\begin{equation}\label{bb}
\left|v[(\hat{B}_1 \hat{B}_{2 n}) \circ (\hat{A}_{1 n} \hat{A}_{2
n})] - v[\hat{B}_1] \, v[ \hat{B}_{2 n}] \, v[\hat{A}_{1 n}] \, v[
\hat{A}_{2 n}]\right| < \epsilon.
\end{equation}
Because of the commutation/anticommutation relations (\ref{commn},
\ref{antin}),
\begin{equation}
(\hat{B}_1\hat{B}_{2 n}) \circ (\hat{A}_{1 n} \hat{A}_{2 n})
 = - (\hat{B}_1 \circ \hat{A}_{2 n})\: \hat{B}_{2 n} \hat{A}_{1 n},
\end{equation}
and therefore,
\begin{equation}\label{cc}
v[(\hat{B}_1\hat{B}_{2 n}) \circ (\hat{A}_{1 n} \hat{A}_{2 n})]
 = - v[(\hat{B}_1 \circ \hat{A}_{2 n})\: \hat{B}_{2 n} \hat{A}_{1
 n}].
\end{equation}
We have, for any $\epsilon > 0$, for sufficiently large $n$,
probability greater than $1-\delta$ that (\ref{aa}) holds, and
probability greater than $1-\delta$ that (\ref{bb}) holds.  Since
$\delta$ was chosen to be less than $\half$, this entails that
there is a nonzero probability that both (\ref{aa}) and (\ref{bb})
hold; in fact, since $\delta$ can be chose arbitrarily small, this
probability can be made arbitrarily close to unity. We therefore
have the conclusion that for any $\epsilon$, for sufficiently
large $n$ the probability is greater than $ 1 - \delta$ that each
of (\ref{aa}) and (\ref{bb}) hold.  Each choice of $n$ will yield
a different set of definite values $\{ v[\hat{B}_1], \:
v[\hat{B}_{2 n} ], \: v[\hat{A}_{1 n}], \, v[\hat{A}_{2 n}] \}$.
The probability distributions for these values must mirror the
quantum-mechanical predictions for the outcomes of measurements of
these quantities; hence, by taking $\epsilon$ sufficiently small,
it is possible to make the probability arbitrarily close to unity
that
\begin{equation}\label{eps}
3 \epsilon <   \left|v[\hat{B}_1]\, v[\hat{B}_{2 n} ] \,
v[\hat{A}_{1 n}] \, v[\hat{A}_{2 n}]\right|.
\end{equation}
Let us assume that we have chosen some $\epsilon$, $n$ such that
this is the case. Let
\begin{eqnarray}
\nonumber x &=& v[(\hat{B}_1\hat{B}_{2 n}) \circ (\hat{A}_{1 n}
\hat{A}_{2 n})] \\ \nonumber y &=& v[(\hat{B}_1 \circ \hat{A}_{2
n})\: \hat{B}_{2 n} \hat{A}_{1
 n}] \\ z &=& v[\hat{B}_1]\, v[\hat{B}_{2 n} ] \,
v[\hat{A}_{1 n}] \, v[\hat{A}_{2 n}].
\end{eqnarray}
Equations (\ref{aa}) and (\ref{bb}) require that $x$ and $y$ both
be a distance less than $\epsilon$ from $z$, and hence a distance
less than $2\epsilon$ from each other.  But $\epsilon$ was chosen
to be smaller than $|z|/3$,  and so $x$ and $y$ must both have
absolute value greater than $2 \epsilon$.  By (\ref{cc}), $x = -
y$, and so their distance from each other must be greater than $4
\epsilon$, contradicting our previous conclusion that the distance
between them is less than $2 \epsilon$. Therefore, (\ref{aa}),
(\ref{bb}), (\ref{cc}), and (\ref{eps}) cannot simultaneously be
satisfied.

\section{Comment}  On the basis of an analogy with
the case of a single spin-$\half$ particle, Clifton \cite{clif}
conjectured that no Kochen-Specker obstruction for position and
momentum using only one degree of freedom was possible.  Since
Clifton's conjecture concerns obstructions as usually conceived,
and not $\epsilon$-obstructions, the $\epsilon$-obstruction in
this paper does not refute this conjecture, which remains
undecided. The $\epsilon$-obstruction in this paper does, however,
reveal a disanalogy with the spin-$\half$ case; since there is, in
fact, a noncontextual hidden-variables theory for single
spin-$\half$ particles that satisfies the $\epsilon$-Product Rule,
there can be no $\epsilon$-obstruction for such a case.



\end{document}